
\input harvmac
\noblackbox
\input epsf
\ifx\epsfbox\UnDeFiNeD\message{(NO epsf.tex, FIGURES WILL BE
IGNORED)}
\def\figin#1{\vskip2in}
\else\message{(FIGURES WILL BE INCLUDED)}\def\figin#1{#1}\fi
\def\ifig#1#2#3{\xdef#1{fig.~\the\figno}
\goodbreak\topinsert\figin{\centerline{#3}}%
\smallskip\centerline{\vbox{\baselineskip12pt
\advance\hsize by -1truein\noindent{\bf Fig.~\the\figno:} #2}}
\bigskip\endinsert\global\advance\figno by1}
%
%
\def\ss{\sigma}
\def\xm{x^-}
\def\xp{x^+}
\def\xv{\vec x}

\def\Xv{\vec X}

\def\pm{p^-}
\def\pp{p^+}
\def\pv{\vec p}

\def\al{\alpha}
\def\nv{\vec n}

%
%

\def\NP{{\it Nucl. Phys.\ }}

\def\PL{{\it Phys. Lett.\ }}
\def\PR{{\it Phys. Rev.\ }}
\def\PRL{{\it Phys. Rev. Lett.\ }}

\def\PRep{{\it Phys. Rep.\ }}

\baselineskip 12pt
\Title{\vbox{\baselineskip12pt \hbox{hep-th/9402136}
\hbox{UCSBTH-94-07} \hbox{SU-ITP-94-5}}}
{\vbox{\hbox{\centerline{Information Spreading in}}
\hbox{\centerline{Interacting
String Field Theory}}}}
\centerline{\it David A. Lowe\foot{lowe@tpau.physics.ucsb.edu}}
\centerline{Department of Physics}
\centerline{University of California}
\centerline{Santa Barbara, CA 93106-9530}
\smallskip
\centerline{\it Leonard
Susskind\foot{susskind@dormouse.stanford.edu}}
\centerline{and}
\centerline{\it John Uglum\foot{john@dormouse.stanford.edu}}
\centerline{Department of Physics}
\centerline{Stanford University}
\centerline{Stanford, CA 94305}
\vskip .5cm
\noindent
The commutator of string fields is considered in the context of light
cone string field theory.  It is shown that the commutator is in
general non--vanishing outside the string light cone.  This could
have profound implications for our understanding of the localization
of information in quantum gravity.

\Date{February, 1994}

\newsec{Introduction}

\nref\wood{D.~A.~Eliezer and R.~P.~Woodard, \NP {\bf B325} (1989)
389.}
\nref\mand{S.~Mandelstam, \NP {\bf B64} (1973) 205; \NP {\bf B83}
(1974) 413; \PRep {\bf 13C} (1974) 259.}
\nref\crem{E.~Cremmer and J.-L.~Gervais, \NP {\bf B76} (1974) 209;
\NP {\bf B90} (1975) 410.}%
\nref\lcstring{M.~Kaku and K.~Kikkawa, \PR {\bf D10} (1974) 1110;
\PR {\bf D10} (1974) 1823.}%
\nref\hop{J.~F.~L.~Hopkinson, R.~W.~Tucker and P.~A.~Collins,
\PR {\bf D12} (1975) 1653.}%
\nref\david{D.~A.~Lowe, ``Causal Properties of Free String Field
Theory,''
UCSBTH-93-39(revised), December 1993, hep-th/9312107, to appear in
\PL B.}
\nref\mart{E.~Martinec, {\it Class. Quant. Grav.} {\bf 10} (1993)
L187; "Strings and Causality'', preprint EFI-93-65, November 1993,
hepth/9311129.}
\nref\lenny{L.~Susskind, ``String Theory and the Principle of Black
Hole Complementarity'',  \PRL {\bf 71} (1993) 2367;
``Strings, Black Holes, and Lorentz
Contraction'', Stanford University preprint SU-ITP-93-21, August
1993, hep-th/9308139.}

Conventional string theory provides us with a perturbative
$S-$matrix.  However, it is of interest to compute more general
observable quantities, such as probability amplitudes at finite
times. In this paper, we will be primarily interested in the causal
properties of such amplitudes, and we will use string field theory to
try and answer these questions.

In the known covariant formulations of quantized string field theory,
the interactions are nonlocal in the center of mass coordinate
$x^{\mu}$, and, as emphasized in \refs{\wood}, the initial value
problem breaks down.  Therefore, the theory cannot be canonically
quantized in the conventional way. Fortunately, one is able to fix
light cone gauge, where the interactions are local in the center of
mass coordinate $x^+ = \tau$. This allows a conventional canonical
quantization, and a second--quantized operator formulation exists
\refs{\mand{--}\hop}, which allows one to perform the kind of
calculations referred to in the preceding paragraph.

Our main concern will be the calculation of the commutator of two
string fields on a flat spacetime background.  In the free theory,
this commutator vanishes outside the ``string light cone''
\refs{\david, \mart}, but it will be shown in this paper that this is
no longer the case when interactions are included.

The plan of the paper is as follows. We begin by reviewing some basic
facts about second--quantized light cone string field theory
\refs{\mand{--}\hop}.  Then we consider the commutator of two string
fields in coordinate representation, and prove that it does not
vanish identically outside the string light cone.  To elucidate the
behavior of this commutator, we calculate a matrix element of the
commutator of tachyon component fields, and show that it is
exponentially damped as the transverse separation increases.  For
certain kinematical configurations, the matrix element can be made
large far outside the light cone.  However, the rate of oscillation
in light cone time of the matrix element is computed, and is shown to
be large in these situations.  The conclusion is that measurements
could detect the information carried by a string state outside the
light cone of the center of mass, but only if these measurements
could be performed with resolution times smaller than the string
scale.  This result supports recent arguments by one of the authors
concerning the nature of information in string theory \refs{\lenny},
and could have profound implications for our understanding of
localization of information in a theory of quantum gravity.

\newsec{Calculation of Commutator}

Let us introduce the light-cone coordinates
\eqn\lcoords{
X^+ = (X^0 + X^{D-1})/ \sqrt{2}, \qquad X^-= (X^0-X^{D-1})/ \sqrt{2}
}
and parametrize the worldsheet of the string by the variables
$\sigma$ and $\tau$.  Light-cone gauge corresponds to fixing
$X^+(\sigma) = \xp=\tau$.  In the following we will consider open
bosonic strings; the generalization to closed strings and to
superstrings should be similar.  The transverse coordinates are
expanded as
\eqn\xtrans{
\vec X(\sigma) = \xv + 2 \sum_{l=1}^{\infty} \xv_l \cos(l \ss)~.
}
In light-cone gauge, the string field is a physical observable and
can be decomposed in terms of an infinite number of component fields.
 In the absence of interactions, the string field takes the form
\refs{\crem{--}\lcstring}
\eqn\pdecom{
\eqalign{\Phi(\tau, \xm, \Xv(\sigma)) = \int {{d^{D-2}p} \over
{(2\pi)^{D-1}}} \int_0^{\infty} {{d\pp} \over {2\pp}} &\biggl [
\sum_{ \{\nv_l \} } A(\pp, \pv, \{ \nv_l \} )
e^{i(\pv \cdot \xv -\pp \xm - \tau \pm)} f_{ \{ \nv_l \} }( \xv_l)
\cr
&+ h.c. \biggr ] \>. \cr }
}
Here the light-cone energy of a string state is given by
\eqn\lcenergy{
p^- \bigl(\pp, \pv, \{ \nv_l \}\bigr) =
{{\pv^{\; 2} + 2\sum_{l, i} l n_l^i + m_0^2} \over
{2\pp}} \>,
}
where $m_0^2$ is the mass squared of the ground state of the string.
For bosonic strings, the ground state is a tachyon and $m_0^2$ is
negative.  In order to effect the light cone quantization and
calculate the commutator, we will regard $m_0^2$ as a positive
adjustable parameter \refs{\david, \mart}.  Of course, this is
inconsistent with Lorentz invariance for the bosonic string, but our
results will be essentially unchanged in the superstring
case where $m_0^2 = 0$.

The $f_{ \{ \nv_l \} }( \xv_l)$ are harmonic oscillator wave
functions given by
\eqn\harmon{
f_{ \{ \nv_l \} }( \xv_l) = \prod_{l=1}^{\infty} \prod_{i=1}^{D-2}
H_{n_l^i}(x_l^i)
e^{-l (x_l^i)^2/ (4\pi) }~,
}
with $H_{n_l^i}(x_l^i)$ a Hermite polynomial.  The $A$ operators obey
the canonical commutation relations
\eqn\commut{
[ A(\pp, \pv, \{ \nv_l \} ), A^{\dag}( {\pp}', {\pv}\,', \{ \nv_l \,'
\} )] = 2 \pp (2\pi)^{D-1} \delta(\pp-{\pp}') \delta^{D-2}(\pv
-{\pv}\,') \delta_{ \{ \nv_l \},
\{ \nv_l\,' \} }~,
}

A component field is obtained from $\Phi$ by multiplying by the
appropriate wave function \harmon\ and integrating over the normal
mode coordinates $\xv_l$.  For example, the tachyon field is given by
\eqn\tacky{
T(\tau, \xm, \xv) = \int {{d^{D-2}p} \over {(2\pi)^{D-1}}} \int
{{d\pp} \over {2\pp}} [ a_T(\pv, \pp) e^{i[\pv \cdot \xv - \pp \xm -
\pm \tau]} + a_T^{\dag}(\pv, \pp) e^{-i[\pv \cdot \xv - \pp \xm -
\pm \tau]}] \>,
}
where $a_T(\pv, \pp) = A(\pv, \pp, \{ {\vec 0} \} )$
and $\pm$ is given by \lcenergy .

Now we want to include a cubic interaction.  The light cone
Hamiltonian becomes $H = H_0 + H_3$, where $H_0$ is the Hamiltonian
for free string field theory and the cubic interaction term $H_3$ is
given by
\eqn\hthree{
\eqalign{
H_3 &= g \int \Phi_{\al_1}(\vec X_1(\ss)) \Phi_{\al_2}(\vec X_2(\ss))
\Phi_{\al_3}(\vec X_3(\ss)) \delta(\sum_{r=1}^3 \al_r)
\Delta( \vec X_1(\ss) - \vec X_2(\ss) - \vec X_3(\ss) )
\cr & \times \mu(\al_1,\al_2,\al_3) \prod_{r=1}^3 d \al_r
\prod_{r=1}^3 {\cal D}
\vec X_r(\ss) \>, \cr}
}
where $g$ is the open string coupling, $\al_r = 2\pp_r$, and the
measure factor is
\eqn\measure{
\mu(\al_1,\al_2,\al_3) = ({\rm det} \Gamma)^{(D-2)/2} \exp (-{{\tau_0
m_0^2} \over 2}
\sum_{r=1}^3
{1\over {\al_r}} ) \>.
}
The infinite-dimensional matrix $\Gamma$ is defined by
\eqn\gamdef{
\eqalign{
\Gamma &= \sum_{r=1}^3 A^{(r)} A^{(r)T}~, \cr
A^{(1)}_{mn} &= \delta_{mn}~, \cr
A^{(2)}_{mn} &= - {2\over {\pi}} \sqrt{mn} (-1)^{m} {{(\beta+1)
\sin(m \pi \beta) } \over { n^2 -m^2(\beta +1)^2 }}~, \cr
A^{(3)}_{mn} &= - {2\over {\pi}} \sqrt{mn} (-1)^{m+n} {{\beta
\sin(m \pi \beta) } \over { n^2 -m^2\beta^2 }}~, \cr
}
}
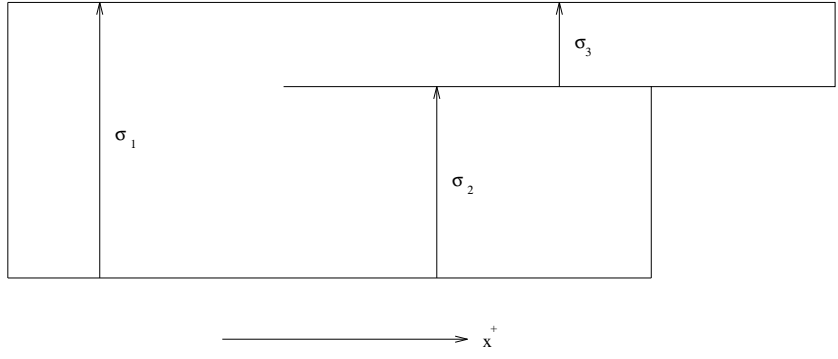
\ifig\fone{Three string interaction.}
{\epsfysize=1.5in\epsfbox{scone.eps}}
\noindent with $\beta = \al_3 / \al_1$ and $\tau_0 = \sum_{r=1}^3
\al_r \log|\al_r|$.  This interaction corresponds to the splitting of
one string into two, as shown in the light-cone diagram \fone .

Now that interactions have been included, we wish to determine
whether the commutator of two string fields vanishes when the
arguments of the string fields lie outside the string light cone
\refs{\david, \mart}.  Suppose that
\eqn\proofone{
[\Phi(\xp_1, \xm_1, \Xv_1(\ss )),
\Phi(\xp_2, \xm_2, \Xv_2(\ss ))] = 0
}
when
\eqn\prooftwo{
{1 \over \pi} \int d\ss (X_1(\ss) - X_2(\ss))^2 < 0 \>,
}
where we are using the mostly minus convention for the
spacetime metric.
For fixed $X_2(\ss)$, equation \proofone\ can be regarded as a
function of $X_1(\ss)$, which vanishes in the entire region in which
equation \prooftwo\ is satisfied.  Differentiating equation
\proofone\ with respect to $\xp_1$ and setting $\xp_1 = \xp_2 =
\tau$, one obtains
\eqn\proofthree{
[\dot \Phi(\tau, \xm_1, \Xv_1(\ss )),
\Phi(\tau, \xm_2, \Xv_2(\ss ))] = 0
}
in the region in which equation \prooftwo\ holds.  Here $\dot \Phi$
denotes $\partial\Phi/ \partial \xp$.  At equal light cone times,
equation \prooftwo\ reduces to
\eqn\prooffour{
{1 \over \pi} \int d\ss (\Xv_1(\ss) - \Xv_2(\ss))^2 > 0 \>.
}
Therefore, to prove that the string field commutator does not vanish
identically outside the string light cone, it is sufficient to prove
that equation \proofthree\ fails to hold when equation \prooffour\ is
satisfied.

To proceed, note that we can use the Heisenberg equation of motion to
express the field $\dot \Phi$ as
\eqn\Hem{
\dot \Phi = i[H, \Phi] \>.
}
Consider now the matrix element
\eqn\Matel{
\eqalign{\vev{0| \Phi(\pp_3, &\Xv_3(\ss_3)) \, [ \dot \Phi(\xm_1,
\Xv_1(\ss_1) ) , \Phi(\xm_2, \Xv_2(\ss_2)) ] | 0 } \cr
&= i \vev{0| \Phi(\pp_3, \Xv_3(\ss_3)) \, \bigl[ [H, \Phi(\xm_1,
\Xv_1(\ss_1) ) ], \Phi(\xm_2, \Xv_2(\ss_2)) \bigr] | 0 } \>, \cr}
}
where all fields are evaluated at $\tau = 0$.  Expanding $H = H_0 +
H_3$, the terms involving $H_0$ all vanish by orthogonality.  This is
a reflection of the fact that the commutator does in fact vanish
outside the string light cone in free string field theory
\refs{\david}.  The remaining terms can be expressed as
\eqn\scomm{
\eqalign{
\vev{0| \Phi(\pp_3, \Xv_3(\ss_3)) \, &[ \dot \Phi(\xm_1, \Xv_1(\ss_1)
) ,
\Phi(\xm_2, \Xv_2(\ss_2)) ] | 0 } = {{2ig(2\pi)^{5(D-1)/2}} \over
{(2\pi)^3 \pp_3}}
\int_0^{\infty} {{d \pp_1}\over {2\pp_1} }
\int_0^{\infty} {{d \pp_2}\over {2 \pp_2} } \cr
\biggl( &e^{i(\pp_1 \xm_1 + \pp_2 \xm_2)}
V(2\pp_1, \Xv_1(\ss_1); 2\pp_2, \Xv_2(\ss_2); -2 \pp_3, \Xv_3(\ss_3))
\cr
&-e^{i(\xm_2 \pp_2 - \xm_1 \pp_1)}
V(-2\pp_1, \Xv_1(\ss_1) ; 2 \pp_2, \Xv_2(\ss_2) ; -2 \pp_3,
\Xv_3(\ss_3) )
\cr &- e^{i(\xm_1 \pp_1 - \xm_2 \pp_2)}
V(2\pp_1, \Xv_1(\ss_1) ; -2 \pp_2, \Xv_2(\ss_2) ; -2 \pp_3,
\Xv_3(\ss_3) )
\biggr) \cr}
}
where $V$ is the vertex factor obtained from \hthree.  The three
terms represent the three possible kinematical situations, in which
the center of mass of string 3, 2, or 1 lies between the centers of
mass of the other two, respectively.  The $\pp_1$ integral may be
performed by using the $\delta(\sum_{r=1}^3 \al_r)$ factor. Then one
notes that the functional $\delta$-function in \hthree\ contains a
zero mode piece $\delta^{D-2} ( \sum_{r=1}^3 \al_r \xv_r )$. Using
one of these delta functions, say for the $x^1$ component, allows the
integral over $\pp_2$ to be performed, and sets
\eqn\longitud{
\eqalign{
\al_2 &= - \al_3 s~, \cr
\al_1 &= (s - 1) \al_3, \cr }
}
where
\eqn\longitudtwo{
s = {{(x_3^1 - x_1^1)} \over {(x_2^1 - x_1^1)}} \>.
}

The crucial point to notice is that one is left with a
$(D-3)$-dimensional $\delta$-function requiring the $\xv_r$ to be
collinear, and that each term in \scomm\ has support on a distinct
ordering of the $\xv_r$ on the line connecting them.  We therefore
find that the commutator of two string fields is in general
non--vanishing outside the string light-cone,
$\int d\ss ( \delta X^{\mu} (\ss) )^2 =0$, when interactions are
included. This also implies the commutator is non--vanishing when the
centers of mass of the strings are spacelike separated.  It should be
stressed here that the non--vanishing of the commutator at spacelike
separations has nothing to do with the fact the bosonic string has a
tachyon. The same will be true in the tachyon--free superstring case.

Of more direct physical interest is the analogous calculation for the
component fields.  For simplicity, we will do the calculation for the
tachyon field, though the generalization to an arbitrary mass
eigenstate is straightforward.  Following the previous line of
reasoning, we compute the matrix element
\eqn\tcomm{
\eqalign{
\vev{0| T(\pp_3, \xv_3) \, [ \dot T(\xm_1, \xv_1),& T(\xm_2,\xv_2) ]
|0} = {{i(2\pi)^{5(D-1)/2}g} \over {\pp_3 (2\pi)^3}} \int {{d
\pp_1}\over {2\pp_1} } \int {{d \pp_2}\over {2 \pp_2} } \cr
&\biggl( e^{i(\xm_1 \pp_1 + \xm_2 \pp_2)} V(2\pp_1, \xv_1; 2\pp_2,
\xv_2; -2\pp_3, \xv_3) \cr
&-e^{i(\xm_2 \pp_2 - \xm_1 \pp_1)} V(-2\pp_1, \xv_1; 2\pp_2, \xv_2;
-2\pp_3, \xv_3)
\cr &- e^{i(\xm_1 \pp_1 - \xm_2 \pp_2)} V(2\pp_1, \xv_1; -2\pp_2,
\xv_2; -2\pp_3, \xv_3)
\biggr) \>,\cr}
}
where, as before, all operators are at time $\tau = 0$.  The vertex
appearing in equation \tcomm\ is the Mandelstam vertex \refs{\mand},
which has the momentum space representation
\eqn\vertex{
V(\al_r, \pv_r) = \delta^{D-2}(\sum_{r=1}^3 \pv_r)
\delta(\sum_{r=1}^3 \al_r)
\exp\biggl ( {{\tau_0} \over 2} \sum_{r=1}^3 { {\pv_r^{\; 2} + m_0^2}
\over {\al_r}} \biggr ) \>.
}
Fourier transforming to coordinate representation, one obtains
\eqn\cvertex{
\eqalign{
V(\al_r, \xv_r) &= \delta^{D-2}(\sum_{r=1}^3 \al_r \xv_r)
\delta(\sum_{r=1}^3 \al_r)
\biggl( {{ \al_1 \al_2 \al_3} \over {8\pi^3  \tau_0}}
\biggr)^{(D-2)/2}
\cr & \times
\exp\biggl( {{\tau_0 m_0^2} \over 2} \sum {1\over {\al_r}} + { {\al_1
\al_2 \al_3}
\over
{8 \tau_0}} ({{\xv_1 -\xv_2}\over{\al_3}} - {{\xv_1-\xv_3}\over
{\al_2}})^2
\biggr) \>. \cr}
}
As was the case for the general string field vertex, equation
\cvertex\ contains the factor $\delta^{D-2}( \sum_{r=1}^3 \al_r
\xv_r)$, so the result is non--vanishing only when the points $\xv_r$
are collinear. The off-shell vertex corresponds to, say, one tachyon
splitting into two others such that all transverse centers of mass
lie along the same line at equal times. In addition there is a
Gaussian factor depending on the separation of the particles.

Consider a configuration in which $\xv_3$ lies between $\xv_1$ and
$\xv_2$, so that only the first term in equation \tcomm\ is
non--zero.  A simple calculation then gives
\eqn\tcomans{
\eqalign{
\vev{0| T(\pp_3, &\xv_3) \, [ \dot T(\xm_1, \xv_1) , T(\xm_2,\xv_2) ]
|0} =
-i e^{i \pp_3 ((1-s) \xm_1 + s \xm_2)} {{\delta^{D-3} (\xv_3 - \xv_1
- s(\xv_2 - \xv_1))} \over {8 \sqrt{2\pi} (\pp_3)^2 s (1-s) |x_2^1 -
x_1^1|}} \cr
&\times \biggl ( {{(2\pi)^2 s (1-s)} \over {\gamma(s)}} \biggr
)^{(D-2)/2}
\exp\biggl( - {{s(1-s)} \over {2 \gamma(s)}} (\xv_1 -\xv_2)^2 -
m_0^2 {{\gamma(s) (s^2-s+1)}\over {2s(1-s)}} \biggr) \>, \cr}
}
where $s$ is given in equation \longitudtwo\ and
\eqn\gamdef{
\gamma(s) = -[s \log (s) + (1-s) \log (1-s)] \>.
}
Note that because of our choice of configuration, $s \in [0, 1]$, and
that $\gamma$ is non--negative.  The matrix element \tcomans\ depends
on the transverse displacement $|\xv_1 -\xv_2|$ through a Gaussian
factor with variance
\eqn\spread{
\sigma^2 = {{\gamma (s)} \over {s(1-s)}} \>.
}
One therefore finds that the matrix element has support over a
distance of order $\sigma^2$ outside the light--cone of the center of
mass.  This spread can be made quite large.  Indeed, for small $s$,
we have
\eqn\limone{
\lim_{s \rightarrow 0} {{\gamma (s)} \over {s(1-s)}} \sim -\log (s)
\>,
}
so for $s \sim \exp \bigl ( -(\xv_1 - \xv_2)^2 \bigr )$, the matrix
element is appreciable.  This can always be achieved by choosing
$x_3^1$ sufficiently close to $x_1^1$.

The question is whether one is able to resolve this information
in practice.
To get an estimate of how quickly the matrix element is
oscillating in light cone time, we can calculate the matrix element
\eqn\tcomdd{
\vev{0| T(\pp_3, \xv_3) \, [\ddot T(\xm_1, \xv_1), T(\xm_2,\xv_2)]|0}
\>,
}
and divide by the matrix element \tcomm .  This is proportional to
the frequency of oscillation.  To do this carefully, we must multiply
both \tcomm\ and \tcomdd\ by a slowly varying function $f$ and then
integrate over $\xv_2, \ldots, \xv_{D-2}$ to eliminate the
$\delta^{D-3}(\sum \al_r \xv_r)$ factors.  Performing this
calculation leads to the following oscillation time scale
\eqn\tres{
\delta t \sim {{\pp_3 s} \over {(\xv_2-\xv_1)^2 \log^{-2} (s) +
(D - 2 + m_0^2)}}
}
valid for $s \rightarrow 0$.  For the case of interest, $s = \exp
\bigl ( -(\xv_1 - \xv_2)^2 \bigr )$, this becomes
\eqn\trestwo{
\delta t \sim {{\pp_3 \exp \bigl ( -(\xv_1 - \xv_2)^2 \bigr )} \over
{(\xv_1 - \xv_2)^{-2} + (D - 2 + m_0^2)}} \>.
}
The conclusion is that in order to observe the spread of information
over more than a string length, one must perform measurements
involving time scales much smaller than the string time.

\newsec{Discussion}

\nref\pmt{A.~Mezhlumian, A.~Peet, and L.~Thorlacius, ``String
Thermalization Near a Black Hole Horizon'', preprint SU-ITP-94-4,
NSF-ITP-94-17, February 1994, hep-th/9402125.}

Two questions concerning the results of the previous section should
be addressed.  The first concerns the meaning of the result \tcomans
{}.
The fact that the commutator is not identically zero for $(\xv_1 -
\xv_2)^2$ spacelike does not signify a breakdown of causality in
string theory, but is merely the result of trying to describe
extended objects by local fields.  Indeed, one should expect that
much of the information carried by a string state is outside the
light cone of the center of mass.  This does not necessarily
mean that signals can propagate faster than the speed of light.

To illustrate what is meant here, consider the case of large $N$ QCD
without matter.  In the $N \rightarrow \infty$ limit, the theory
describes noninteracting, extended objects, namely glueballs.  One
could choose to write down a free field theory for the glueballs,
using operators which create and annihilate glueballs, and these
fields will commute when their arguments are spacelike separated.
Once $N$ is allowed to be finite, however, interactions must be
included.  If one continues to use the glueball fields, one will find
that the interactions appear highly nonlocal, and fields will fail to
commute when their arguments are spacelike separated.  Despite all
this, we know that causality is not violated, because the underlying
theory is QCD, which is a causal, local quantum field theory.

Quantum electrodynamics provides another example.  If one chooses to
perform calculations in Coulomb gauge, one finds the commutator of
the gauge field is non--vanishing at spacelike separations.  This is
because there are only two physical degrees of freedom (the fields
transverse to the direction of propagation of a photon, for example)
and the third field must be expressed as a nonlocal function of the
other two, using the gauge constraint.  When one computes the
commutator of gauge invariant quantities such as the electromagnetic
field strength, however, the apparent acausality disappears.  It is
the nonlocality in the description of the $A$ field that causes this
problem.

Both of the above examples differ from string field theory in that
for both cases there exists an underlying causal, local field theory.
 The above examples simply show that if one chooses to describe
extended objects in these theories by local fields, one must be
careful in interpreting what will appear to be acausal results.
String field theory is different because the only known formulations
of string field theory are inherently nonlocal.  This is not to say
that there does not exist a local formulation of string field
theory--this remains an open question.  There is, however, a fairly
large amount of evidence showing that such a local formulation does
not exist \refs{\wood}, and we believe that this is probably the
case.  In any event, because strings are extended objects, it is
incorrect to conclude from the above results that causality is
violated in string theory.  It should also be noted that the
non--vanishing of the commutator is a result of the interactions
between strings.  As was shown in \refs{\david, \mart}, free string
fields do commute outside the string light cone.  This is also
evidenced by the fact that the matrix element \tcomans\ is
proportional to the open string coupling $g$.

The second point to be addressed is the question of Lorentz
covariance of the expression \tcomans .  Although Lorentz covariance
is not manifest, this does not mean that the matrix element \tcomans\
is not
Lorentz covariant (for $m_0^2 = -2$).  If it were, this would
contradict the statement
made previously that there is no violation of causality in string
field theory.  To see why, remember that in order to even define the
light cone string field theory, one must fix a light cone gauge.  The
fields then depend explicitly on this choice of coordinates.  In
order to leave light cone gauge, one must understand what the light
cone fields are in terms of the degrees of freedom of a covariant
string field theory.  In general, the light cone fields will be
complicated functionals of the new degrees of freedom, and certainly
should not be expected to have simple Lorentz transformation
properties.  Said another way, the light cone string fields are
defined as functionals of loops which have a particular orientation
in spacetime.  When one makes a Lorentz transformation, the loops on
which the new string fields are defined are a different set of loops
with a different orientation, and the Lorentz transformation
properties must include terms which transform one set of loops to
another.  The transformation will be very complicated.

The pathologies of the bosonic string introduce some ambiguity into
the interpretation of our results.  We again stress that the
calculation of the commutator of bosonic string fields presented here
can be done in the tachyon free superstring case, and one will obtain
a similar answer.

The information content of these matrix elements exhibits precisely
the same type of diffusive behavior as was described in
\refs{\lenny}, which was argued to provide a possible resolution of
the black hole information paradox.  Under conditions relevant to
strings propagating near a horizon, the spread of the matrix elements
\tcomm\ can become arbitrarily large.  As an explicit example, we can
replace the field $T(\tau, \xm_2, \xv_2)$ in equation \tcomm\ by
$T(\tau, \pp_2, \xv_2)$, and treat this field as representing a
string which falls toward the horizon of a large black hole of mass
$M$.  If the string starts off with light cone momentum $\pp$ at
Schwarzschild time $t=0$, then after a time $t$ has elapsed, the
momenta will be given by $\pp_2 \sim \pp e^{-t/4M}$.  The field
$T(\tau, \xm_1, \xv_1)$ can be interpreted as a string close to the
horizon at a fixed position, and $T(\tau, \pp_3, \xv_3)$ can be
interpreted as a test string which remains far from the horizon.
Setting $\pp_3 = \pp$, the matrix element contains the same Gaussian
factor as equation \tcomans , with $s = e^{-t/4M}$.  One thus finds
that the spread of the Gaussian is given by $\sigma^2 \sim {t \over
{4M}}$.  This diffusive behavior is the same as that found in
\refs{\lenny, \pmt}.

\bigskip
{\bf Acknowledgements:}

The authors would like to thank A.~Peet for helpful discussions.
D.~L. was supported in part by National Science Foundation grant
PHY-91-16964.  L.~S. was supported in part by National Science
Foundation grant PHY89-17438.  J.~U. was supported in part by
National Science Foundation grant PHY89-17438 and by a National
Science Foundation Graduate Fellowship.

\vfill\eject
\listrefs
\end